\documentclass[12pt]{article}
\begin{document}
\title{The Dark Matter Puzzle And Other Issues}
\author{B.G. Sidharth\\
International Institute for Applicable Mathematics \& Information Sciences\\
Hyderabad (India) \& Udine (Italy)\\
B.M. Birla Science Centre, Adarsh Nagar, Hyderabad - 500 063
(India)}
\date{}
\maketitle
\begin{abstract}
We consider the problem of the flattening of the velocity curves in
galactic discs and the consequent postulation of dark matter from
three different but converging perspectives-- a change in the large
scale dimensionality of space, a  variation of $G$ and the MOND
approach. We also discuss the paradigm of the universe itself being
a Black Hole.
\end{abstract}
\section{Introduction}
It is many decades now since the existence of dark, that is non
luminous matter was postulated, though the identity of this dark
matter has only been a matter of guess work. The reason for invoking
the hypothetical dark matter is well known-- the velocities of stars
in galaxies should tend to zero using usual dynamics, as we approach
the edge of the disc. Instead astrophysical observation has
consistently shown that the velocity curves flatten out, that is the
velocities tend a constant rather than zero. So Zwicky and others
postulated that there was matter other than the visible matter which
gave a greater mass to the galaxies, and this in turn would explain
the velocity discrepancy \cite{nar}.\\
The question then arose, "What exactly is this dark matter?". Over
the years several hypotheses have been put forth-- it could be hot
dark matter or it could be cold dark matter. These could range from
weakly interacting massive particles (WIMPS) to cold neutrinos. Or
there could be the missing monopoles, or undetectable brown dwarf
stars or even black holes and so on. To this day the question has
remained unresolved.\\
There have however been alternative suggestions to explain the flat
velocity curves. We will discuss three of these, two put forward by
the author and the other by Milgrom, and try to find an
interrelationship.
\section{Less Than Three Dimensional Space?}
The author (with A.D. Popova) \cite{popova} suggested that the
dimensionality of space falls off asymptotically, and this would
explain astrophysical observations including the dark matter
problem. Indeed it had already been argued that the dimensionality
of space could be expressed by a non integer number that is less
than three on large scales \cite{pop2,pop3}. Indeed the three
dimensionality of our immediate space may be necessary, for the very
existence of atoms, as was pointed by Ehrenfest long ago
\cite{ehrenfest}. Similarly this dimensionality may also be required
for usual wave propagation \cite{barrow,cu,uof}. All this is at what
we may call intermediate scales. At different scales of measurement,
the dimensionality could be different \cite{csf}. This fact could
explain the dark matter problem, as we will now argue.\\
More generally, the dependence of matter on distance $M(r)$,
obtained from observing $21 cm$ neutral hydrogen emission of gas
clouds moving around a galaxy far from its visual bounds (the
continuation of a rotation curve) \cite{bur,kor} is
\begin{equation}
M \propto r^{1.2 \div 1.3}\label{e1}
\end{equation}
This conclusion reflects the fact that the observed rotation
velocity slightly increases at outer parts of galaxies, so the
growth of $M$ is interpreted as the presence of some dark halo
besides luminous matter. Moreover, the amount of dark matter grows
relatively to luminous matter when coming to larger and larger
scales \cite{davies}.\\
However, even the nonrelativistic (Newtonian) consideration of
gravitational forces in spaces with lesser than three dimensions,
enables us in principle to bring in correlation dynamics and "the
shortage" of luminous matter. It would be very difficult to take
into account the smooth fall of dimensionality because we do not
know the law of such a fall. In order to make some estimates we
roughly assume that on some relative distance $R_0$ the
dimensionality changes by a leap from $3$ to $n < 3$. We consider
the rotation curves of disk galaxies --similar considerations apply
for the dynamics of double galaxies and the dispersion of velocities
in elliptical (spheroidal) galaxies. We show how one can lower the
estimates of masses of these systems under the assumption that the
dimensionality is lesser than three starting even at scales of the
order of a typical galaxy's size. We also discuss the possible
hierarchical change of dimensionality. We only demonstrate the
principal possibility to lower dynamical mass and refrain from
making any numerical estimates as yet because of
too many uncertain parameters.\\
Certainly, we know now of a constructive physical model which can
describe noninteger and nonconstant dimensionality. We can only
outline some suggestive arguments for it. The first of possible
suggestions comes from the fractal theory \cite{turner}; the space
itself may have a fractal-like structure. The second suggestion is
that effectively, if we consider individually each object in the
Universe leaving aside other objects, then we can perceive a space
between us and this object as $2$-dimensional because one spatial
direction is fixed as a line from us to the object, and the other
direction can be fixed by a vector of relative velocity of the
object with respect to us. Thus, may be our space filled by distant
separated objects consists of a set of (perhaps non-connected)
$2$-dimensional subspaces for which effectively $2 \leq n < 3$. The
third suggestion comes from the existence of the large-scale
structure of the Universe in distribution of galaxies, their groups,
clusters, voids and superclusters. There is the tendency for matter
to perform oblate structures at each hierarchical level. Possibly,
the structure and dimensionality of our space itself might reflect
the distribution of (luminous) cosmic matter.
\section{Newtonian Consideration}
In $3$ and $n$ dimensions, the expressions for the gravitational
forces acting between the mass $M$ and a unit mass separated by the
distance $r$ are
\begin{equation}
F^{(3)} = -\frac{G^{(3)} M}{r^2}\label{e2.1}
\end{equation}
and
\begin{equation}
F^{(n)} = - \frac{G^{(n)} M}{r^{n-1}}\, ,\label{e2.2}
\end{equation}
respectively, where $G^{(3)}$ and $G^{(n)}$ are relevant
gravitational constants. The corresponding potentials $(\vec{F} = -
\vec{\nabla}\Phi$ by definition) up to arbitrary constants $C^{(3)}$
and $C^{(n)}$ are
\begin{equation}
\Phi^{(3)} = -\frac{G^{(3)} M}{r} + C^{(3)},\label{e2.3}
\end{equation}
and for $n \ne 2$
\begin{equation}
\Phi^{(n)} = -\frac{1}{n-2} \frac{G^{(n)} M}{r^{n-2}} +
C^{(n)}\label{e2.4}
\end{equation}
In this case $n = 2$,
$$\Phi^{(2)} = G^{(2)} M lnr + C^{(2)}.$$
Now, let us assume that at some relative distance $R_0$ from a body
the dimensionality changes by leap from $3$ to $n$. The condition of
matching the forces (\ref{e2.1}) and (\ref{e2.2}) at $R_0$ gives the
connection between the gravitational constants
\begin{equation}
G^{(n)} = G^{(3)} R^{n-3}_0\label{e2.5}
\end{equation}
Thus, the improved force in accordance with our conception is
(\ref{e2.1}) for $r \leq R_0$ and (\ref{e2.2}) for $r > R_0$ with
(\ref{e2.3}):
\begin{equation}
F^{imp} = \left\{\begin{array}{ll} F^{(3)}, & r \leq R_0,\\
F^{(n)}, & r > R_0.
\end{array}\right.
\label{e2.6}
\end{equation}
The condition of matching the potentials (\ref{e2.3}) and
(\ref{e2.4}) at $R_0$ is also required, and leads to the following
expression for the "n-dimensional" constant $C^{(3)} = 0$ is chosen
in (\ref{e2.3}),
$$C^{(n)} = \frac{3-n}{n-2} \frac{G^{(3)} M}{R_0}$$
for $n \ne 2$, and
$$C^{(2)} = - \frac{G^{(3)} M}{R_0} (1 + ln R_0)$$
for $n = 2$.\\
Thus, the improved potential in accordance with our conception is
(\ref{e2.3}) for $r \leq R_0$ and (\ref{e2.4}) for $r > R_0$:
\begin{equation}
\Phi^{imp} = \left\{\begin{array}{ll} \Phi^{(3)}, & r \leq R_0,\\
\Phi^{(n)}, & r > R_0.
\end{array}\right.
\label{e2.7}
\end{equation}
Let us stress that we can only think of
$R_0$ as of a relative distance between any bodies. Otherwise, first
the conception of relativity of space which is an achievement of
Einstein's physics, would be violated. Second, there would be
troubles with the
universality of a gravitational attraction.\\
Indeed, let a sphere with the radius $R_0$ be "rigidly fixed" in
space, and let bodies freely come in it and come out of it. Let a
body of the mass $M$ be placed inside the sphere at some point $O$
which does not coincide with the center of the sphere $O$, and let
other bodies of the equal small mass $m$ be outside of this sphere
at the points $A_1$ and $A_2$ in such a way that $O'A_1 = O'A_2 =
r$. Then, we have for the gravitational forces between each body of
the mass $m$ and that of the mass $M$
$$F^{(n)}_1 = - G^{(n)}_1 \frac{mM}{r^{n-1}}, F^{(n)}_2
\frac{mM}{r^{n-1}}$$ where $G^{(n)}_1 = G^{(3)} R_1^{n-3}$ and
$G^{(n)}_2 = G^{(3)} R^{n-3}_2$ are the gravitational constants
found from the continuity of forces at the sphere, and $R_1$ and
$R_2$ are the distances from the point $O'$ to the points of
intersections of the rays $O'A_1$ and $O'A_2$ with the sphere. Thus,
if the above consideration were true, then the "gravitational
constant" would not be a constant, and the law of gravitation
attraction would be anisotropic and in-homogeneous. Moreover, in the
case $n < 3$, the body of the mass $M$ at $O'$ which tends to the
sphere on the insider $(R_1 \to 0 \, \mbox{or}\, R_2 \to 0)$ would
produce an infinitely divergent "gravitational constant".\\
That is why we admit $R_0$ as only relative distance. Perhaps, our
consideration would be less rough, if we consider the change of
dimensionality which occurs by leaps several times from $3$ to $n_1$
at $R_0$, from $n_1$ to $n_2$ at $R_1$, and so on, and from $n_j$ to
$n_{j+1} > 2 \, \mbox{at}\, R_j$. Then, we have the chain of
relations between the gravitational constants
$$G^{(n_1)} = G^{(3)} R^{n_1-3}_0$$
\begin{equation}
G^{(n_2)} = G^{(n_1)} R^{n_2-n_1}_1 = G^{(3)} R^{n_1-3}_0
R^{n_2-n_1}_1\label{e2.8}
\end{equation}
$$G^{(n_j+1)} = G^{(n_j)} R^{n_{j+1}^{1-n_j}}_j = \cdots =
G^{(3)}R^{n_1-3}_0 R^{n_2-n_1}_1 \cdots R^{n_{j+1}^{-n_j}}_j$$ The
chain of relations between the constants in potentials is rather
cumbersome; however the recurrence relation is
$$C^{(n_{j+1})} = G^{(n_j)} R^{n_{j+1}^{-n_j}}_j = \cdots = G^{(3)}
R^{n_1-3}_0 R^{n_2-n_1}_1 \cdots R^{n_{j+1}-n_j}_j.$$
The chain of
relations between the constants in potentials is rather cumbersome;
however the recurrence relation is
$$C^{(n_{j+1})} = C^{(n_j)} + \frac{n_j -
n_{j+1}}{(n_j-2)(n_{j+1}-2)} \frac{G^{(n_j)} M}{R^{n_j-2}_j}$$ The
three subsections below present the application of the force
(\ref{e2.6}) (the potential of (\ref{e2.7}) to determinations of
galactic masses. We shall return to discussing the problem of the
dimensionality change after considering the three examples below.
\section{Rotation Curves of Disk Galaxies}
A rough calculation of rotation velocity $v$ at the distance $r$ far
from the center of a galaxy (i.e., when all its mass is effectively
concentrated near the center or spherically distributed around it)
is based on the equality of the centrifugal force and gravitational
force.\\
In the $3$-dimensional space, using (\ref{e2.1})
\begin{equation}
\frac{v^2}{r} = \frac{G^{(3)} M^{(3)}_g}{r^2}\label{e2.9}
\end{equation}
In accordance with our conception, we can write
$$M^{dyn}_g \equiv M^{(3)}_g = \frac{v^2r}{G^{(3)}},$$
i.e., we call the dynamical galactic mass, $M^{dyn}_g$, a mass
calculated as if our space were $3$-dimensional.\\
In the $n$-dimensional space, if $r \gg R_0$ then equality
(\ref{e2.9}) should be replaced by the following (with the use of
(\ref{e2.2}):
$$\frac{v^2}{r} = \frac{G^{(n)}M^{(n)}_g}{r^{n-1}},$$
so that we call a mass calculated in the $n$-dimensional space the
true galactic mass, $M^{true}_{g}$:
\begin{equation}
M^{true}_g \equiv M^{(n)}_g = \frac{v^2r^{n-2}}{G^{(n)}} =
\left(\frac{R_0}{r}\right)^{3-n} M^{dyn}_g\label{e2.10}
\end{equation}
where in the last equality (\ref{e2.5}) is used. When $2 < n < 3$
the factor at $M^{dyn}_g$ is lesser than unity, therefore
$M^{true}_g < M^{dyn}_g$.\\
The more accurate calculation of the rotation curve can be done for
the case when the disk of a galaxy lies in the $2$-dimensional space
(or plane), and there exist no other spatial dimensions. Let the
distribution of the $2$-dimensional matter density in the disk
satisfy the law
\begin{equation}
\rho = \rho_0 exp \left(\frac{r}{R_d}\right)\label{e2.11}
\end{equation}
where $\rho_0$ is the $2$-density in the disk center, and $R_d$ is
some characteristic radius. Let $R_0 \ll R_d$. The distribution
(\ref{e2.11}) corresponds to the observed distribution of luminous
matter in \cite{turner}. The velocity square in this case is given
as follows
\begin{equation}
v^2(r) = 2 \pi \rho_0 G^{(3)} \frac{R^2_d}{R_d} \left[1 - \left(1 +
\frac{r}{R_d}\right) exp \left(-
\frac{r}{R_d}\right)\right]\label{e2.12}
\end{equation}
The function (\ref{e2.12}) monotonically increases from zero and
tends to a constant value at $r \to \infty$. That is why the
dynamical mass calculated with the aid of (\ref{e2.12}) tends to
grow linearly at larger $r:M^{dyn} \propto r$. However, at any
finite $r$ we can effectively write $M^{dyn}_g \propto r^\beta$
where always $\beta > 1$. Probably, this fact could explain the
dependence (\ref{e1}), meaning that our real space has the
dimensionality which is very near to two at the scales of the outer
parts of galaxies. This also explains the flattening of the rotational curves,
without invoking dark matter.\\
Alternatively, we note that from the above, for $w = 2$, we get
$$v^2 = GMlnr,$$
or
$$v \frac{dv}{dr} \propto \frac{1}{r} \to 0$$
as $r$ becomes large, so that $\frac{dv}{dr} \to 0$, because $v$
does not $\to 0$. So, $v \to$ a constant value.
\section{The Time Variation of the Gravitational Constant}
We now come to the author's cosmological model which in 1997
predicted a dark energy driven accelerating universe with a small
cosmological constant. It may be recalled that at that time the
ruling paradigm embodied in the hot big bang standard cosmological
model was exactly the opposite. This model has been discussed in
detail (Cf.ref.\cite{cu,uof,ijmpa,ijtp}). In this model it turns out
that the gravitational constant, rather as in Dirac cosmology, has
the following time dependence
\begin{equation}
G = \frac{\beta}{T}\label{e14}
\end{equation}
where in our cosmology $\beta$ is given in terms of the constant
microphysical parameters.\\
We next observe that from (\ref{e14}) it follows that
\begin{equation}
G = G_0 \left(1 - \frac{t}{t_0}\right)\label{e15}
\end{equation}
where $G_0$ is the present value of $G$ and $t_0$ is the present age
of the universe and $t$ the time elapsed from the present epoch.
Similarly one could deduce that (Cf.ref.\cite{nar}),
\begin{equation}
r = r_0 \left(\frac{t_0}{t_0 + t}\right)\label{e16}
\end{equation}
We next use Kepler's Third law \cite{gold}:
\begin{equation}
\tau = \frac{2 \pi a^{3/2}}{\sqrt{GM}}\label{e17}
\end{equation}
$\tau$ is the period of revolution, $a$ is the orbit's semi major
axis, and $M$ is the mass of the sun. Denoting the average angular
velocity of the planet by
$$\dot \Theta \equiv \frac{2 \pi}{\tau},$$
it follows from (\ref{e15}), (\ref{e16}) and (\ref{e17}) that
$$\dot \Theta - \dot \Theta_o =  \dot \Theta_0 \frac{t}{t_o},$$
where the subscript $o$ refers to the present epoch, \\
Whence,
\begin{equation}
\omega (t) \equiv \Theta - \Theta_o =  \frac{\pi}{\tau_o t_o}
t^2\label{e18}
\end{equation}
Equation (\ref{e18}) gives the average perhelion precession at time
'$t$'. Specializing to the case of Mercury, where $\tau_o =
\frac{1}{4}$ year, it follows from (\ref{e18}) that the average
precession per year at time '$t$' is given by
\begin{equation}
\omega (t) =  \frac{4\pi t^2}{t_0}\label{e19}
\end{equation}
Whence, considering $\omega (t)$ for years $t=1,2, \cdots , 100,$ we
can obtain from (\ref{e19}), the usual total perhelion precession
per century as,
$$\omega = \sum^{100}_{n=1} \omega (n) \approx 43'' ,$$
if the age of the universe is taken to be $\approx 2 \times 10^{10}$ years.\\
Conversely, if we use the observed value of the precession in
(\ref{e19}),
we can get back the above age of the universe.\\
Interestingly it can be seen from (\ref{e19}), that the precession depends on the epoch.\\
We next demonstrate that orbiting objects will have an anamolous
inward
radial acceleration.\\
Using the well known equation for Keplarian orbits
(cf.ref.\cite{gold}),
\begin{equation}
\frac{1}{r} = \frac{GMm^2}{l^2} (1 + e cos \Theta)\label{e20}
\end{equation}
\begin{equation}
\dot r^2 \approx \frac{GM}{r} - \frac{l^2}{m^2r^2}\label{e21}
\end{equation}
$l$ being the orbital angular momentum constant and $e$ the
eccentricity of the orbit, we can deduce such an extra inward radial
acceleration, on differentiation of (\ref{e21}) and using
(\ref{e15}) and (\ref{e16}),
\begin{equation}
a_r = \frac{GM}{2t_o r \dot r}\label{e22}
\end{equation}
It can be easily shown from (\ref{e20}) that (on the average),
\begin{equation}
\dot r \approx \frac{eGM}{rv}\label{e23}
\end{equation}
For a nearly circular orbit $rv^2 \approx GM$, whence use of
(\ref{e23}) in (\ref{e22}) gives,
\begin{equation}
a_r \approx v/2 t_o e\label{e24}
\end{equation}
For the earth, (\ref{e24}) gives an anomalous inward radial
acceleration
$\sim 10^{-9} cm/sec^2,$ which is known to be the case \cite{kuhne}.\\
We could also deduce a progressive decrease in the eccentricity of
orbits. Indeed, $e$ in (\ref{e20}) is given by
$$e^2 = 1+\frac{2El^2}{G^2m^3M^2} \equiv 1 + \gamma , \gamma < 0.$$
Use of (\ref{e15}) in the above and differenciation, leads to,
$$\dot e = \frac{\gamma}{et_o} \approx - \frac{1}{et_o} \approx - \frac{10^{-10}}
{e} \mbox{per year},$$ if the orbit is nearly circular. (Variations
of eccentricity in the usual theory have been extensively studied
(cf.ref.\cite{berger} for a review). On the other hand, for open
orbits, $\gamma > 0,$ the eccentricity would
progressively increase.\\
We finally consider the anomalous accelerations given in (\ref{e22})
and (\ref{e24})
in the context of space crafts leaving the solar system.\\
If in (\ref{e22}) we use the fact that $\dot r \leq v$ and
approximate
$$v \approx \sqrt{\frac{GM}{r}},$$
we get,
$$a_r \geq \frac{1}{et_o} \sqrt{\frac{GM}{r}}$$
For $r \sim 10^{14}cm$, as is the case of Pioneer $10$ or Pioneer
$11$, this gives,
$a_r \geq 10^{-11}cm/sec^2$\\
Interestingly Anderson et al.,\cite{anderson} claim to have observed
an anomalous inward acceleration
of $\sim 10^{-9} cm/sec^2$.\\
\section{Other Consequences}
We could also explain the correct gravitational bending of light.
Infact in Newtonian theory also we obtain the bending of light,
though the amount is half that predicted by General
Relativity\cite{denman,silverman,brill}. In the Newtonian theory we
can obtain the bending from the well known orbital equations
(Cf.also(\ref{e20})),
\begin{equation}
\frac{1}{r} = \frac{GM}{L^2} (1+ecos\Theta)\label{e25}
\end{equation}
where $M$ is the mass of the central object, $L$ is the angular
momentum per unit mass, which in our case is $bc$, $b$ being the
impact parameter or minimum approach distance of light to the
object, and $e$ the eccentricity of the trajectory is given by
\begin{equation}
e^2 = 1+ \frac{c^2L^2}{G^2M^2}\label{e26}
\end{equation}
For the deflection of light $\alpha$, if we substitute $r = \pm
\infty$, and then use (\ref{e26}) we get
\begin{equation}
\alpha = \frac{2GM}{bc^2}\label{e27}
\end{equation}
This is half the General
Relativistic value.\\
We next note that the effect of time variation of $r$ is given by
equation (\ref{e16})(cf.ref.\cite{nc115}). Using (\ref{e16}) the
well known equation for the trajectory is given by
(Cf.\cite{gold,berg,lass})
\begin{equation}
u" + u = \frac{GM}{L^2} + u\frac{t}{t_0} + 0 \left (
\frac{t}{t_0}\right )^2\label{e28}
\end{equation}
where $u = \frac{1}{r}$ and primes denote differenciation with
respect to
$\Theta$.\\
The first term on the right hand side represents the Newtonian
contribution while the remaining terms are the contributions due to
(\ref{e16}). The solution of (\ref{e28}) is given by
\begin{equation}
u = \frac{GM}{L^2} \left[ 1 + ecos\left\{
\left(1-\frac{t}{2t_0}\right ) \Theta +
\omega\right\}\right]\label{e29}
\end{equation}
where $\omega$ is a constant of integration. Corresponding to
$-\infty < r < \infty$ in the Newtonian case we have in the present
case, $-t_0 < t < t_0$, where $t_0$ is large and infinite for
practical purposes. Accordingly the analogue of the reception of
light for the observer, viz., $r = + \infty$ in the Newtonian case
is obtained by taking $t = t_0$ in (\ref{e29}) which gives
\begin{equation}
u = \frac{GM}{L^2} + ecos \left(\frac{\Theta}{2} + \omega
\right)\label{e30}
\end{equation}
Comparison of (\ref{e30}) with the Newtonian solution obtained by
neglecting terms $\sim t/t_0$ in equations (\ref{e16}),(\ref{e28})
and (\ref{e29}) shows that the Newtonian $\Theta$ is replaced by
$\frac{\Theta}{2}$, whence the deflection obtained by equating the
left side of (\ref{e30}) to zero, is
\begin{equation}
cos \Theta \left(1-\frac{t}{2t_0}\right) = -\frac{1}{e}\label{e31}
\end{equation}
where $e$ is given by (\ref{e26}). The value of the deflection from
(\ref{e31}) is twice the Newtonian deflection given by (\ref{e27}).
That is the deflection $\alpha$ is now given not by (\ref{e28}) but
by the formula,
\begin{equation}
\alpha = \frac{4GM}{bc^2},\label{e32}
\end{equation}
The relation (\ref{e32}) is the correct observed value and is the
same as
the General Relativistic formula.\\
We now come to the problem of galactic rotational curves
(cf.ref.\cite{nar}). We would expect, on the basis of
straightforward dynamics that the rotational velocities at the edges
of galaxies would fall off according to
\begin{equation}
v^2 \approx \frac{GM}{r}\label{e33}
\end{equation}
However it is found that the velocities tend to a constant value,
\begin{equation}
v \sim 300km/sec\label{e34}
\end{equation}
This has lead to the postulation of as yet undetected additional
matter, the so called dark matter.(However for an alternative view
point Cf.\cite{sivaram}. We observe that from (\ref{e16}) it can be
easily deduced that\cite{edge}
\begin{equation}
a \equiv (\ddot{r}_{o} - \ddot{r}) \approx \frac{1}{t_o}
(t\ddot{r_o} + 2\dot r_o) \approx -2 \frac{r_o}{t^2_o}\label{e35}
\end{equation}
as we are considering infinitesimal intervals $t$ and nearly
circular orbits. Equation (\ref{e35}) shows (Cf.ref\cite{nc115}
also) that there is an anomalous inward acceleration, as if there is
an extra attractive force, or an additional central mass, as indeed
we saw a little earlier.\\
So,
\begin{equation}
\frac{GMm}{r^2} + \frac{2mr}{t^2_o} \approx
\frac{mv^2}{r}\label{e36}
\end{equation}
From (\ref{e36}) it follows that
\begin{equation}
v \approx \left(\frac{2r^2}{t^2_o} + \frac{GM}{r}\right)^{1/2}
\label{e37}
\end{equation}
From (\ref{e37}) it is easily seen that at distances within the edge
of a typical galaxy, that is $r < 10^{23}cms$ the equation
(\ref{e33}) holds but as we reach the edge and beyond, that is for
$r \geq 10^{24}cms$ we have $v \sim 10^7 cms$
per second, in agreement with (\ref{e34}).\\
Thus the time variation of G explains observation
without invoking dark matter.\\
\section{The MOND Approach}
Milgrom \cite{milgrom} approached the problem by modifying Newtonian
dynamics at large distances. This approach is purely
phenomenological. The idea was that perhaps standard Newtonian
dynamics works at the scale of the solar system but at galactic
scales involving much larger distances perhaps the situation is
difference. However a simple modification of the distance dependence
in the gravitation law, as pointed by Milgrom would not do, even if
it produced the asymptotically flat rotation curves of galaxies.
Such a law would predict the wrong form of the mass velocity
relation. So Milgrom suggested the following modification to
Newtonian dynamics: A test particle at a distance $r$ from a large
mass $M$ is subject to the acceleration $a$ given by
\begin{equation}
a^2/a_0 = MGr^{-2},\label{em1}
\end{equation}
where $a_0$ is an acceleration such that standard Newtonian dynamics
is a good approximation only for accelerations much larger than
$a_0$. The above equation however would be true when $a$ is much
less than $a_0$. Both the statements can be combined in the
heuristic relation
\begin{equation}
\mu (a/a_0) a = MGr^{-2}\label{em2}
\end{equation}
In (\ref{em2}) $\mu(x) \approx 1$ when $x >> 1, \, \mbox{and}\,
\mu(x) \approx x$ when $x << 1$. It must be stressed that
(\ref{em1}) or (\ref{em2}) are not deduced from theory, but rather
are an ad hoc fit to explain observations. Interestingly it must be
mentioned that most of the implications of MOND do not
depend strongly on the exact form of $\mu$.\\
It can then be shown that the problem of galactic velocities is
solved \cite{mil1,mil2,mil3,mil4,mil5}.
\section{Interrelationship}
It is interesting to note that there is interesting relationship
between the varying $G$ approach, which has a theoretical base and
the purely phenomenological MOND approach. Let us write
$$\beta \frac{GM}{r} = \frac{r^2}{t_0^2} \, \mbox{or} \, \beta =
\frac{r^3}{GMt_0^2}$$
Whence
$$\alpha_0 = v^2/r = \frac{GM}{r^2} \, \alpha = \frac{r}{t_0^2}$$
So that
$$\frac{\alpha}{\alpha_0} = \frac{r^3}{GMt_0^2} = \beta$$
At this stage we can see a similarity with MOND. For if $\beta << 1$
we are with the usual Newtonian dynamics and if $\beta > 1$ then we
get back to the varying $G$ case exactly as with MOND.\\
Furthermore, as can be seen from (\ref{e2.5}), when the
dimensionality $n$ gets smaller than $3$, effectively $G$ starts
falling off as in the time varying case seen in sections $5$ and
$6$.
\section{Discussion}
It is interesting to note that the varying $G$ approach leads to
several observations that have been carried out including the
precession of the perihelion of the planets, in particular Mercury,
the Pioneer anomaly, the shortening of the orbital periods of binary
pulsars and so on \cite{nc115,varyG,cu,uof}.\\
A further interesting observation is the fact noted by Milgrom that
there is a curious coincidence in MOND viz.,
\begin{equation}
a_0 \sim H (\sim 10^{-7} cm sec^{-2})\label{ex}
\end{equation}
where $H$ is the Hubble constant. In fact this follows from the
varying $G$ theory. For, we have in this case from (\ref{e22}),
(\ref{e23}), (\ref{e24}) and (\ref{e37}),
$$a_0 \sim r/t_0^2$$
Feeding the values of $r$, the radius of the universe $= ct_0$ and
the fact that $H \sim \frac{1}{t_0}$, we get (\ref{ex}), which now
shows up no longer as a coincidence.\\
Finally, it may be noted that the Boomerang results are in tune with
MOND rather than the Dark Matter scenario, the WMAP model
notwithstanding \cite{gauge}.\\
It is also interesting to note that for large $r$, (\ref{e37}) gives
the Hubble law, here deduced from the $G$ variation.
\section{Higher Dimensionality}
The fact that asymptotically the universe appears to be two
dimensional is interesting from a totally different point of view.
This is that a collection of ultra relativistic masses would appear
as two dimensional as discussed in detail in ref.\cite{moller}. This
is what may be called an inside view of the universe. We now come to
what may be called the outside view. The starting point is the
observation that the universe shows up as a black hole. This is
because we have,
\begin{equation}
R \sim \frac{GM}{c^2}\label{eB}
\end{equation}
Equation (\ref{eB}) is not phenomenological-- indeed it can be
deduced from theory as discussed in detail (Cf.ref.\cite{uof}). The
equation describes a Schwarzchild black hole. In (\ref{eB}), dark
matter if any is not included. If however dark matter were included
in the mass then the case for a black hole becomes even stronger, as the left
scales becomes smaller than the right side.\\
We now argue that this three dimensional black hole, which describes
our universe is imbedded in a four dimensional manifold as in
Wheeler's Superspace or in the author's multiple universe model
\cite{mcu}. We now extend the Lorentz transformation to a four
dimensional space. In this case the angular momentum
$$\int \vec{r} \times \vec{p} dV$$
is replaced by
$$\vec{\omega} = \int \vec{r} \Lambda \vec{p} dV$$
where $\Lambda$ represents the usual anti-symmetrical product.\\
Whence we have
\begin{equation}
\vec{r} \cdot \vec{\omega} = 0\label{ez}
\end{equation}
The significance of (\ref{ez}) is that it represents a three
dimensional hyper surface in four dimensional space, remembering
that $\vec{r}$ has four space components. This apart the two
dimensional surface in the ultra relativistic case referred to above
follows precisely from (\ref{ez}), but this time in three
dimensional space. To sum up what this shows is that indeed the
three dimensional universe can be considered to be a black hole
imbedded in four dimensional space. There is a totally different
route to the above consideration. As shown by the author and Popova
\cite{popova}, (Cf. also sections 3 and 4), we have
\begin{equation}
\dot{\epsilon} + n \frac{\dot{a}}{a} \epsilon = 0\label{e3.8}
\end{equation}
where $a$ is the size of the universe and $\epsilon$ the energy, and
$n$ the number of dimensions. An integration gives
\begin{equation}
\epsilon \sim Mc^2 a^{-n}\label{e3.9}
\end{equation}
From here it follows that
\begin{equation}
R \equiv a = \sqrt{\frac{(n-1)c^2}{K^{(n)}\epsilon}}\label{eA}
\end{equation}
where
\begin{equation}
K^{(n)} = \frac{n-1}{n-2} \cdot \frac{2\pi^{n/2}}{\Gamma (n/2)}
\cdot \frac{G^{(n)}}{c^2}\label{eC}
\end{equation}
where $G^{(n)}$ has been defined earlier. Substitution of (\ref{eC})
in (\ref{eA}) for the case $n = 4$ gives us back (\ref{eB}) as
required for a black hole. The same conclusion can be drawn
alternatively from similar reasoning \cite{deds} in which case we
get
$$a^{(n)}_{max} = [0(1) \cdot \frac{G^n M}{c^2}]^{\frac{1}{n-2}}$$
which gives back
$$R (\sim 10^{28}cm) \sim \frac{GM}{c^2}$$
for the case $n = 4$.\\
Interestingly if the universe were treated as a black hole, we could
associate with it a spin given by
$$\bar{h} = MRc \sim 10^{93}$$
As pointed out elsewhere \cite{cu} this value seems to be vindicated
by COBE observations and of the order of the spin in Godel's
solution of Einstein's equations. Equally interesting is the fact
that with the scaled up value of the spin $\bar{h}$, the universe
could be considered to be a wave packet with the "Compton
wavelength"
$$R = \frac{\bar{h}}{Mc}$$
To put it another way (Cf.\cite{uof}) the universe could be a wave
packet of a scaled up Schrodinger equation.\\
In any case, we had argued that dimensionality increases with scale
(\cite{uof}) -- from one at the Planck scale to two at the Compton
scale, through three at our scale and now four and beyond at scales
large than $R$ above (Cf. also \cite{mcu}).\\
We conclude with the following remarks, in the context of some of
the above considerations. As discussed in detail elsewhere (Cf. for
example \cite{uof}) the Virial Theorem gives in the astrophysical
context
$$\frac{GM}{R} \sim v^2$$
where $v$ is the dispersion in velocity, $M$ is the mass and $R$ the
extent of a relatively isolated gravitational system. On the other
hand as we saw above we have
$$\frac{R^2}{M^2R^2} \sim v^2$$
Whence we have
$$G \sim \frac{\bar{h}^2}{M^3 R}$$
which gives the correct value of the gravitational constant, if we
feed in the mass of the universe and its radius and also the above
value of the scaled Planck constant $\bar{h}$. What we have done is,
we have obtained the expression for $v$ from two totally different
considerations. The first was, based on the gravitation of the
matter in this mass collection. The second was based on non
dynamical considerations, namely a generalized Schrodinger equation
and Gaussian wave packets. So the latter considerations show up
gravitation as being non fundamental and being more of an effective
force. Indeed the non fundamental nature of gravitation has been
commented upon in detail in the above reference.\\
The other aspect is that the Hubble law is a consequence of the
spread of the above Gaussian wave packet of the scaled up
Schrodinger equation. Indeed all this including the varying $G$ was
related to a dark energy driven universe in the author's work.\\
The interesting point is that the above considerations apply at
other scales also as discussed in the reference. For example at the
scale of galaxies we have a scaled up Planck constant $\sim
10^{74}$. If we now carry over the Gauss packet considerations to
this scale, the decay of the Gauss packet which is the counterpart
of the Hubble expansion of the universe now comes up as the gradual
increase in size of the galaxies. (Of course we can also deduce the
correct value of the universal constant of gravitation in this case
too by similar considerations).

\end{document}